\documentclass[iop,revtex4]{emulateapj}
\usepackage{color}
\usepackage{ulem}

\shorttitle{The GRO~J1655$-$40 wind}
\shortauthors{Shidatsu et al.}

\begin{document}

\title{An optically-thick disk wind in GRO J1655$-$40?}

\author{M. Shidatsu\altaffilmark{1}, C. Done\altaffilmark{2}, Y. Ueda\altaffilmark{3}}
\email{megumi.shidatsu@riken.jp}

\altaffiltext{1}{Institute of Physical and Chemical Research (RIKEN), 2-1 Hirosawa, 
Wako, Saitama 351-0198, Japan}
\altaffiltext{2}{Department of Physics, University of Durham, South Road, Durham, DH1 3LE, UK}
\altaffiltext{3}{Department of Astronomy, Kyoto University, Kitashirakawa-Oiwake-cho, 
Sakyo-ku, Kyoto 606-8502, Japan}

\begin{abstract}
We revisited the unusual wind in GRO J1655$-$40 detected with {\it Chandra} in 
2005 April, using long-term {\it RXTE} X-ray data and simultaneous 
optical/near-infrared photometric data. This wind is the most convincing case 
for magnetic driving in black hole binaries, as it has an inferred launch radius 
which is a factor of 10 smaller than the thermal wind prediction. 
However, the optical and near infrared fluxes monotonically increase around 
the {\it Chandra} observation, whereas the X-ray flux monotonically decreases 
from 10 days beforehand. Yet the optical and near infrared fluxes are from the 
outer, irradiated disk, so for them to increase implies that the X-rays likewise 
increased. We applied a new irradiated disk model to the multi-wavelength 
spectral energy distributions (SEDs). Fitting the optical and near-infrared fluxes, 
we estimated the intrinsic luminosity at the {\it Chandra} epoch was 
$\gtrsim 0.7 L_{\rm Edd}$, which is more than one order of magnitude larger than 
the observed X-ray luminosity. These results could be explained if a Compton-thick, 
almost completely ionized gas was present in the wind and strong scattering reduced 
the apparent X-ray luminosity. 
The effects of scattering in the wind should then be taken into account 
for discussion of the wind-driving mechanism. Radiation pressure and Compton 
heating may also contribute to powering the wind at this high luminosity. 
\end{abstract}

\keywords{accretion, accretion disks --- black hole physics --- 
X-rays: binaries --- X-rays: individual(GRO~J1655$-$40)}

\section{Introduction}
Since the first detection of highly ionized absorption lines in the 
late 1990's, high-resolution X-ray spectroscopy has provided 
great opportunities to probe disk winds in black hole (BH) and 
neutron star (NS) low-mass X-ray binaries \citep[e.g.,][]{ued98, 
kot00, ued01, boi03, dia06, mil06, kub07, mil08, dia14}. 
Disk winds are thought to have an equatorial structure, extending 
along the disk plane with a relatively small solid angle, because 
the lines are only seen in systems with high inclination angles larger 
than $\approx 60^\circ$ \citep{pon12} and do not show large 
orbital dependence of variability \citep{yam01}.
The high mass loss rates, comparable to or even an order 
of magnitude larger than the mass accretion rates 
\citep{ued10, nei11}, suggest that disk winds play a key 
role in the dynamics of accretion onto compact objects, 
although how and how much they affect the accretion disks 
have been poorly understood.

There are three major mechanisms proposed for powering 
the disk winds in low-mass X-ray binaries: radiation pressure, 
thermal pressure (due to Compton heating) and magnetic 
processes. Radiation pressure works effectively at 
near- or super-Eddington luminosity, if gas is highly ionized. 
Thermal winds can be launched from anywhere above a certain 
radius at which the isothermal sound speed of gas heated by 
X-rays overcomes the local escape velocity. This radius, called 
"Compton radius" ($R_{\rm C}$) is given by 
\begin{equation}
R_{\rm C} \approx 10^{10} \left( \frac{M_{\rm BH}}{M_\sun} \right) 
\left( \frac{T_{\rm C}}{10^8 \, \mbox{K}} \right)^{-1} \, \mbox{cm},
\end{equation} 
where $M_{\rm BH}$ and $T_{\rm C}$ are the black hole mass and 
Compton temperature, respectively. More detailed calculations 
suggest that the thermal winds can be launched from smaller 
radii, $\gtrsim$0.1 $R_{\rm C}$ \citep{beg83, woo96}, which 
correspond to $10^4$--$10^5$ $R_{\rm g}$ 
($R_{\rm g}$ is the gravitational radius: $GM_{\rm BH}/c^2$)
in stellar-mass BH X-ray binaries. Ionized gas located within 
$R_{\rm C}$ remains bound to the surface of the accretion 
disks as hot atmosphere. 

Since a wind is photoionized by X-ray radiation, its launching 
radius $R$ can be observationally estimated from the ionization 
parameter $\xi$ defined as, 
\begin{equation}
\xi = \frac{L_{\rm X}}{n R^2}, \label{eq_xiparam}
\end{equation}
where $L_{\rm X}$ and $n$ represent the ionizing 
luminosity and the number density of the wind, respectively. 
Constraining $n$ directly is often not easy, and in such 
cases, the column density of the wind 
($N_{\rm H} \sim n \Delta R$, where $\Delta R$ is the length 
of the wind) has been used under the assumption 
$\Delta R \sim R$, to derive $R$ through the above equation 
\citep[e.g.,][]{ued98, kot00, kub07}.
\citet{dia13} showed that outflowing and static ionized 
absorbers in NS X-ray binaries have $R$ values 
consistent with the thermal driven winds and atmosphere, 
respectively. Many winds detected in BH X-ray binaries are 
also consistent with the thermal and/or 
radiation-pressure driven winds 
\citep[e.g.,][]{kot00, kub07, nei09, dia14}. The wind velocities 
are about 100--1000 km s$^{-1}$, in agreement with the 
prediction that the winds are generated in the outer parts 
of the disks.

In 2005, a remarkable wind from GRO J1655$-$40 
was detected with {\it Chandra} High Energy Transmission 
Grating (HETG) \citep{mil06, net06, mil08}. The observed 
spectrum exhibited a rich forest of blueshifted ionized absorption 
lines. The estimated intrinsic luminosity (0.03 $L_{\rm Edd}$\footnote{$L_{\rm Edd}$ 
(Eddington Luminosity) $= 1.5 \times 10^{38} (M_{\rm BH}/M_\sun)$~erg~s$^{-1}$ 
for the solar abundances.}) is not very high and hence 
the wind cannot be powered by radiation pressure. The thermal 
wind interpretation is also difficult, because this luminosity is just 
about the critical luminosity 
\citep[$L_{\rm crit} \approx 0.03 L_{\rm Edd}$;][]{beg83} 
below which the Compton heating is not efficient enough to drive 
a thermal wind. 

As the result of detailed photoinization modeling, the 
GRO J1655$-$40 wind was found to have an ionization 
parameter of $\xi \sim 10^4$~erg~cm~s$^{-1}$. 
The density was directly estimated to be $n \sim 10^{14}$~cm$^{-3}$ 
from a density-sensitive line ratio of Fe XXII. 
By using Equation~(\ref{eq_xiparam}), 
the radius was derived to be $R = 10^{9.7}$~cm, which is one 
order of magnitude smaller than the radius at which thermal 
driven winds can form. The disk wind was thus explained by 
the remaining scenario, i.e., a magnetic driven wind, although 
the observed outflow velocity ($\approx 1000$~km~s$^{-1}$) 
is lower than the local escape velocity at the radius of 
$\lesssim 10^{9.7}$~cm ($\gtrsim 4000$~km~s$^{-1}$) 
and seems too low to drive a wind,
unless it has a much larger velocity in the perpendicular direction.

Recently, \citet{utt15} have found that unusual properties 
in X-ray short-term variability and continuum spectra 
were seen around the {\it Chandra} observation, from  
a comprehensive study of {\it Rossi X-ray Timing Explorer} 
({\it RXTE}) data, covering almost the entire outburst episode 
in 2005. 
When a deep absorption structure emerged at $\sim 8$ keV, 
the hard X-ray flux was greatly reduced and a very soft 
X-ray spectrum appeared. The peculiar spectral shape was 
similar to that of the "hypersoft state" sometimes observed 
in the high mass X-ray binary Cyg X-3, which has a massive 
stellar wind from its Wolf-Rayet companion \citep[e.g.,][]{kol10}. 
Meanwhile, the variability power above $\sim$0.1~Hz was 
gradually reduced as the observed X-ray flux decreased. 
They suggested that these behaviors are linked to the 
evolution of the disk wind or the accretion 
flow structure, yet the exact relations among them is still 
not completely clear. 

In this article, 
we focus on the shape of the multi-wavelength SED of GRO J1655$-$40 
and its evolution around the {\it Chandra} observation 
during the outburst, and reconsider the wind in the context of the 
long-term behavior of the accretion disk structure. 
We use the {\it RXTE} X-ray data and also simultaneous 
optical and near-infrared (OIR) photometric data from 
the SMARTS (the Small and Medium Aperture Research 
Telescope System). 
The OIR emission can be regarded as a reference of the 
intrinsic luminosity at the outermost part of the accretion disk, 
which would be less affected by reprocessing in the wind. 
The multi-wavelength SEDs thus allow us to connect the 
properties of inner and outer disks. 

In Section~\ref{sec_longterm}, we study the long-term X-ray 
and OIR behavior around the {\it Chandra} observation. 
In Section~\ref{sec_spec}, we analyze the multi-wavelength 
SED at the {\it Chandra} epoch using a new irradiated disk model,
and compare the result with that in the normal high/soft state. 
Compiling all these results, we discuss the nature of the wind 
in Section~\ref{sec_discussion}. We used HEASOFT version 
6.13 to reduce and analyze the data. In the following sections, 
errors represent the 90\% confidence range for a single 
parameter, unless otherwise stated. Throughout the article, 
we refer to the table given by \citet{wil00} as the solar 
abundances. The distance and inclination angle of 
GRO J1655$-$40 are assumed to be $D=3.2$ kpc 
\citep{hje95, oro97} and $i=70^\circ$ \citep{gre01}, 
respectively. 

\section{Long-term trends in optical, near-infrared, X-ray properties} \label{sec_longterm}

\begin{figure}
\plotone{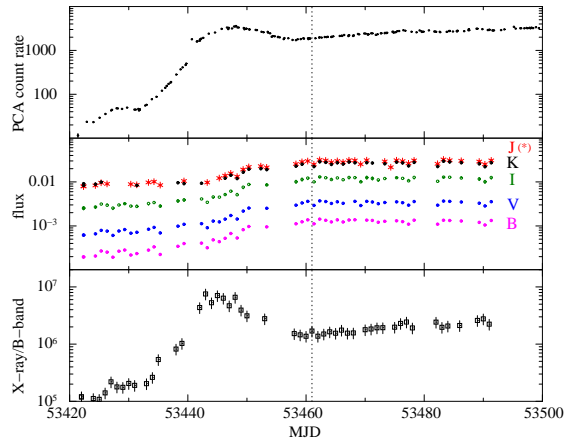}
\caption{Top: Long-term {\it RXTE}/PCA light curve of GRO~J1655$-$40 
in the X-ray band. Middle: Optical and near-infrared light curves obtained 
with SMARTS in $B$ (pink), $V$ (blue), $I$ (green), $J$ (red, marked with 
asterisks), and $K$(black) bands. Bottom: The ratios of the observed B-band 
fluxes with respect to X-ray count rates. The PCA count rates are estimated 
from the Standard 2 data of the Proportional Counter Array 2 (PCU2) in the 
full energy band. The SMARTS data are plotted in units of Jy and not 
corrected for interstellar extinction. 
The dotted line indicates MJD 53461, corresponding to the {\it Chandra} observation. 
\label{fig_lc}}
\end{figure}

Figure~\ref{fig_lc} shows the long-term X-ray and optical/NIR 
behavior in the first $\sim$80 days of the outburst in 2005, 
including the date of the {\it Chandra} observation 
(April 1 $=$MJD 53461). The X-ray count rates are obtained with 
the {\it RXTE}/Proportional Counter Array (PCA), and the OIR fluxes 
are taken from the SMARTS photometric observations in the 
$B$, $V$, $I$, $J$, and $K$ bands 
\citep[see][for the details of the observations]{bux05,mig07}. 

The most remarkable point is that the OIR fluxes were monotonically 
increasing in the whole period, while the X-ray flux was decreasing 
before the {\it Chandra} epoch. This cannot be explained by an 
increase of the mass accretion rate alone. 
As noticed from the bottom panel of Fig.~\ref{fig_lc}, the ratio between 
the X-ray and B-band fluxes marks the minimal value at around the 
{\it Chandra} observation, after the hard-to-soft transition at 
$\sim$MJD 53440.
We note that, among the 5 OIR bands, the $B$-band flux should have 
the least fractional contribution of the blackbody emission from the 
companion (a F6 III-type star with a temperature of $\approx$6300 K; 
\citealt{sha99}), and is dominated by the irradiated disk emission from 
the outer disk (see Section~\ref{sec_spec} for more details).

Using the OIR fluxes and simultaneous {\it RXTE}/Photon Counting 
Array (PCA) and High Energy X-ray Timing Experiment (HEXTE) data, 
we created multi-wavelength SEDs in the individual MJDs. 
We employed the standard data-reduction procedure described in the 
{\it RXTE} cook book and extracted the PCA spectra from the "Standard 2" 
data of the Proportional Counter Array 2 (PCA2) and the HEXTE spectra 
from Cluster A and B data for each observation ID. The OIR magnitudes 
are converted to the fluxes, from which we produced spectra and 
response matrix files in the XSPEC format utilizing the ftool flx2xsp. 
We added a 0.5\% systematic error to each bin in the PCA spectra, 
following previous works \citep[e.g.,][]{mig07, utt15}.

\begin{figure}
\plotone{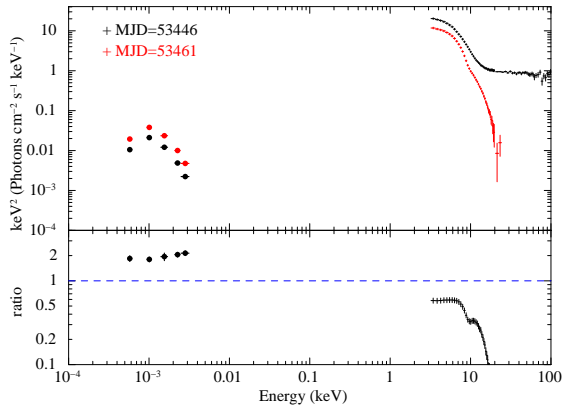} 
\caption{Top: multi-wavelength SEDs in the normal high/soft state 
(MJD 53446, black) and at the {\it Chandra} epoch in the hypersoft state 
(MJD 53461, red). For the HEXTE data (above 20 keV), the Cluster-A 
spectra are not plotted for illustrative purposes. Bottom: the ratio 
of the two SEDs. The latter data are divided by the former.
\label{fig_sed}}
\end{figure}

In Figure~\ref{fig_sed}, the SED on MJD~53461, the date of the 
{\it Chandra} observation, is compared with that on MJD~53446 
in the normal high/soft state, whose X-ray spectrum is well described 
by a multi-color disk (MCD) combined with a hard tail with a photon 
index of $\approx$2. Note that MJD~53446 corresponds to 
the peak of the observed X-ray flux in the initial outburst rise.
This figure shows that the X-ray flux levels decreased at all 
energies from MJD~53446 to the {\it Chandra} epoch, whereas 
the OIR fluxes increased.

Moreover, a huge difference is present in the hard X-ray band.
The difference is not just in normalization. 
The hard tail in the normal high/soft state is suppressed in 
the spectrum on MJD~53461, which is dominated by the soft 
X-ray fluxes below 10~keV.
As reported by \citet{utt15}, this X-ray spectrum can be 
described with a MCD plus an unusually steep power-law 
component with a photon index of $>$5.
This peculiar hypersoft state 
was steadily seen from a few days before the {\it Chandra} 
observation, after making rapid and frequent transitions between 
the normal high/soft and hypersoft states in the preceding 
$\approx$10 days. 
We note that, in Figure~\ref{fig_sed}, a significant depression 
is present at $\approx$8~keV in the SED on MJD 53461, 
which is likely an ionized iron-K edge along with absorption 
lines produced by the disk winds. This feature was always 
visible during the hypersoft state, and deepened without 
significant energy shifts, as the 
observed X-ray flux decreased \citep[see][]{utt15}.

As noticed in the bottom panel of Figure~\ref{fig_sed}, the flux 
ratio of the two observations is almost constant in the soft X-ray 
band below $\sim$5~keV, where the disk blackbody dominates. 
This indicates that the apparent disk luminosity ($L_{\rm disk}$) 
decreased without significantly changing the inner disk 
temperature ($T_{\rm in}$). In fact, the hypersoft state produces 
a vertical branch in the $T_{\rm in}$ vs. $L_{\rm disk}$ diagram 
\citep[see Figure 16 in][]{don07}. This means that the evolution 
of the disk spectrum in the hypersoft state deviates from the 
standard $L_{\rm disk} \propto T_{\rm in}^4$ relation.

What is unusual in the hypersoft state is not only 
the characteristics of the SED but also fast X-ray 
variability. 
As reported by \citet{utt15}, the high-frequency part of 
the flat noise component was remarkably reduced in the 
hypersoft state, compared with that in the normal 
high/soft state. 
The break frequency in the high/soft state was located 
at $\approx$5 Hz, whereas during the hypersoft state, 
it moved to lower frequencies with decreasing 
X-ray flux, to reach $\approx$0.1 Hz at the 
{\it Chandra} observation. 
They also found that the energy dependence of the 
PDS shape was not strong,  
suggesting that the decline of the high frequency 
variability was not simply caused by the 
disappearance of the power-law tail.

\section{A new irradiated disk model "optxrplir" and its application 
to the SEDs} \label{sec_spec}

Here we model the multi-wavelength SEDs of GRO J1655$-$40 
in the normal soft state and at the {\it Chandra} epoch in the 
hypersoft state to test the possibility that the peculiar behaviors 
in the hypersoft state are associated with the disk wind. 
Using the OIR data as a reference of the intrinsic emission from 
the outermost disk region, 
we investigate whether and how much the disk wind can affect 
the intrinsic X-ray emission at the {\it Chandra} epoch. 

\subsection{The {\tt optxrplir} model}

Different spectral components (the companion star, the outer 
parts of the accretion disk, the synchrotron radiation from the jets 
or the hot accretion flow) can contribute to the OIR emission of BH 
X-ray binaries (see e.g., \citealt{van81}, \citealt{gie08}, and 
\citealt{gie09} for contribution from outer disks, \citealt{cor02}, 
\citealt{hyn03}, \citealt{bux04}, \citealt{rus07}, and \citealt{gan10} 
for that from jets, and \citealt{pou14a}, \citealt{pou14b}, and 
\citealt{vel15} for that from hot flows).
Normally, jet emission does not contribute in the disk dominant 
states. In fact, \citet{mig07} found that the radio flux actually 
dropped down to an undetectable level and that jets were 
quenched in the hypersoft state.
Usually, the outer disk emission is the major contributor in 
outbursting BH X-ray binaries with a low-mass companion. 
The outer disk region is illuminated by X-rays from the 
inner disk and the Comptonized corona, and consequently 
the reprocessed emission enhances the OIR fluxes. 
To determine the accurate mass accretion rates through 
the inner and outer disks, we have to account for the 
intrinsic disk, Compton-scattered, and reprocessed 
emission components in a self-consistent manner.

For this purpose, we use a new irradiated disk model named 
"{\tt optxrplir}" and apply it to the multi-wavelength SEDs. This 
model is based on the {\tt optxirr} model \citep{sut14}, which 
combines the two model implemented in XSPEC: "{\tt optxagnf}" 
\citep{don12}, the energetically-coupled disk plus Comptonized 
corona model including color temperature correction, and the 
irradiated disk model "{\tt diskir}" \citep{gie08, gie09}. Using our 
new model, we can fully consider the emissions from the inner 
and outer disk self-consistently. 

The main difference between {\tt optxirr} and our model 
{\tt optxrplir} is the radial dependence of the X-ray flux 
irradiating the disk. 
The {\tt optxrplir} model assumes that the illuminating 
flux does not scale with radius as $r^{-2}$, but $r^{-12/7}$.
The reprocessed component 
in this model is thus more enhanced in outermost parts 
of the disk than the previous models. This radial 
dependence is derived from a more realistic calculation, 
by solving the hydrostatic and energy balance at each 
radius of a geometrically thin irradiated disk \citep{cun76}. 
In this condition, the disk height increases at larger radii in 
proportion to $r^{9/7}$ due to heating by 
irradiation. The $r^{-12/7}$ dependence was recently 
adopted to model multi-wavelength SEDs in the low/hard 
state (\citealt{vel13}). 
Also, previous irradiated disk models describe the 
strength of the reprocessed component with one 
parameter $F_{\rm out}$ (the fraction of 
illuminating flux thermalized in the outer disk), 
while {\tt optxrplir} separates it, for the convenience 
of calculation, into two explicit parameters: 
the geometry-dependent factor $f_{\rm out}$ and 
$(1 - a_{\rm out})$, where $f_{\rm out}$ is 
the height of the disk at the outer edge divided by 
the outer radius and $a_{\rm out}$ is the albedo of 
the outer disk.
We assume that the total X-rays, not only hard X-rays, 
are responsible for reprocessing. This is consistent 
with that the OIR SEDs were not largely changed 
in spite of the huge difference in the hard X-ray tail 
between the high/soft state and the hypersoft state.

The model includes two different Comptonization 
components in the similar way as the 
{\tt optxagnf} model. One is a 
power-law component extending to the hard X-ray band 
above 10 keV, which can be used to model the hard tail 
observed in the normal high/soft state. The other is a 
low-temperature thermal Comptonization component, 
which can be used for the steep power-law component  
sometimes seen at slightly higher energies than a MCD 
component \citep[e.g.,][]{fro01, ibr05, nak12}. We adopt the latter 
to model the steep power-law component in the 
hypersoft state. The former component (hereafter referred 
to as ``power-law tail'') is described with a photon index 
and a cut-off energy, and the latter (``soft thermal 
Comptonization component'') with an electron 
temperature ($kT_{\rm e}$) and an optical depth ($\tau$). 
The {\tt optxrplir} model assumes that the power-law tail 
is generated from the gravitational energy dissipated within 
the radii from the innermost stable circular orbit (ISCO) to 
$R_{\rm pl}$, and that the soft thermal Comptonization 
component is produced between $R_{\rm pl}$ and 
$R_{\rm cor}$. Only disk blackbody emission 
(including irradiation) is emitted from the radii larger 
than $R_{\rm cor}$.
The rest parameters in the model are all taken from the 
{\tt optxagnf} model (the BH mass and spin, the distance 
to the system, and the Eddington ratio) and the {\tt diskir} 
model (the outer disk radius $R_{\rm out}$, 
and also $f_{\rm out}$ and $a_{\rm out}$ as explained above).

\begin{figure*}
\plottwo{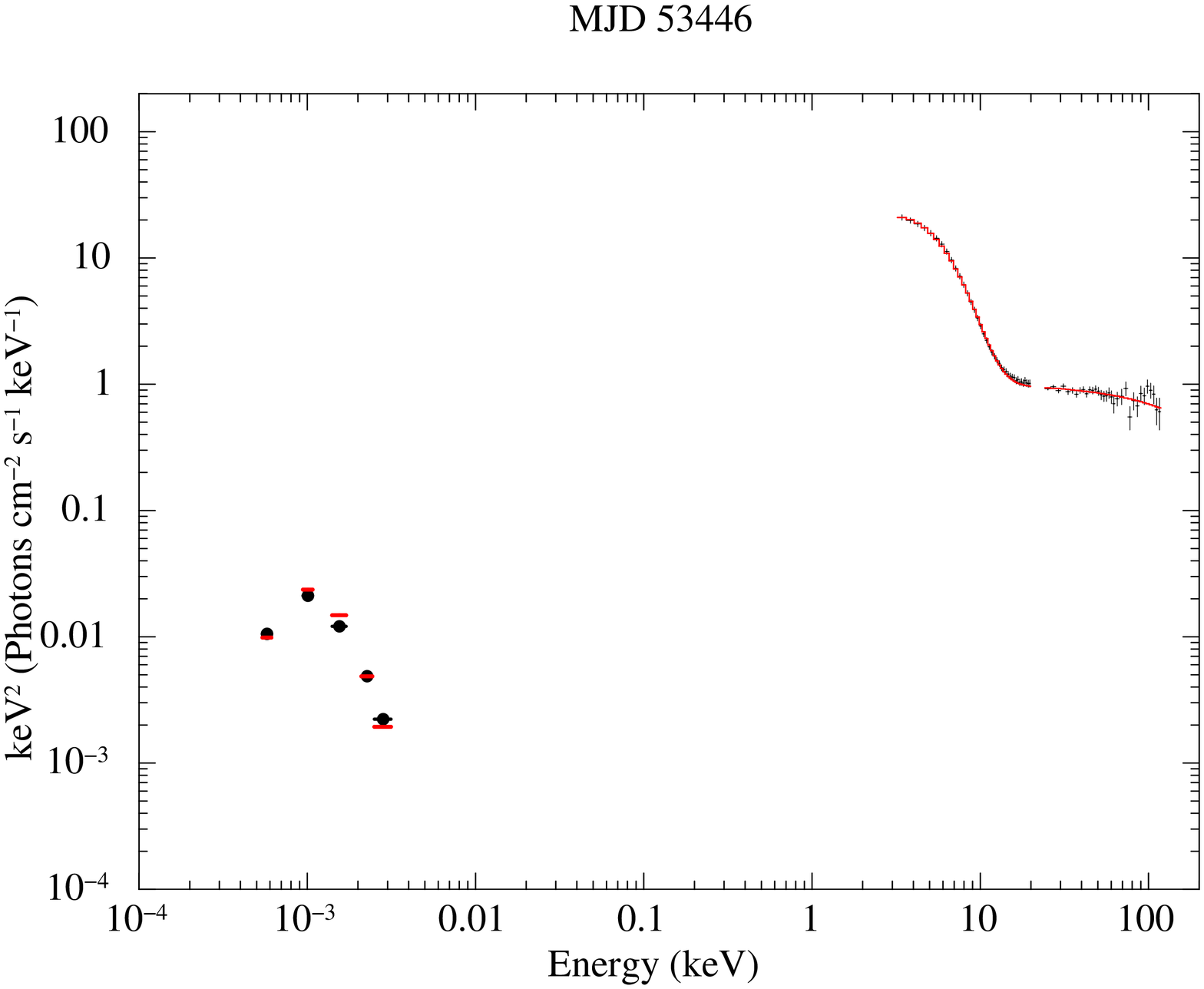}{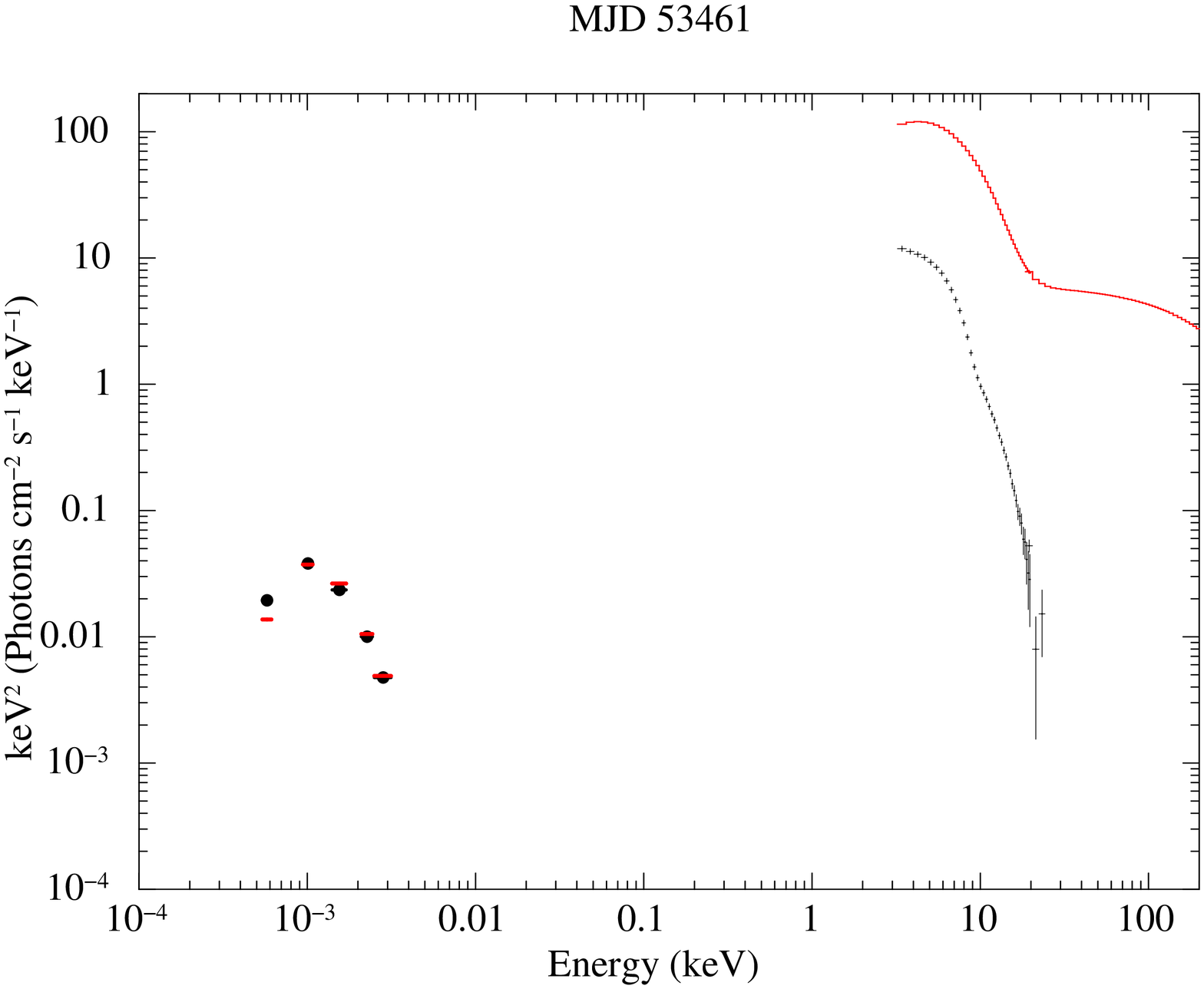}
\caption{Left: the SED and best-fit {\tt optxrplir+bbodyrad} model in the normal 
high/soft state (MJD 53446). Right: the SED at the {\it Chandra} epoch 
(MJD 53461) plotted with the same model except for the luminosity, which 
is adjusted to fit the flux levels in the OIR bands. 
\label{fig_fit}}
\end{figure*}

\begin{figure}
\plotone{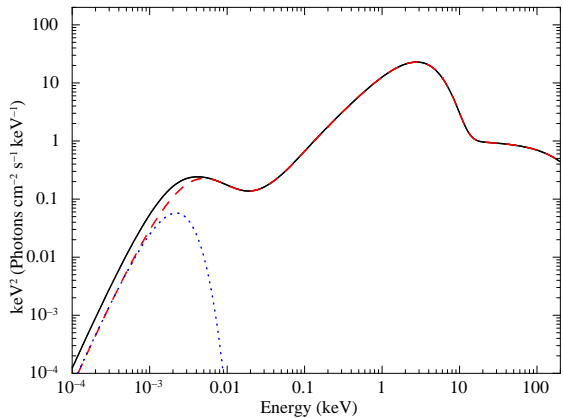}
\caption{The best-fit absorption/extinction-corrected {\tt optxrplir\\+bbodyrad}  
model in the normal high/soft state (MJD 53446). 
The dashed and dotted lines represent the {\tt kerrconv*optxrplir} component 
(including the irradiated disk emission and the hard tail) and the {\tt bbodyrad} 
component (the blackbody emission from the companion star), respectively.
\label{fig_hssmodel}}
\end{figure}

\subsection{SED analysis with {\tt optxrplir}}
\subsubsection{Model setup}
To analyze our multi-wavelength SEDs of GRO J1655$-$40,
we use {\tt optxrplir} as a full model of the irradiated disk 
emission and its Comptonization. The OIR fluxes also 
include the blackbody emission from the companion star, 
for which we use the {\tt bbodyrad} model. 
Throughout the SED analysis, we fixed the temperature 
and radius of the companion star at 6300 K and 
5.0 $R_\sun$, respectively, and the BH mass 
at 6.3 $M_\sun$, following previous estimation by \citet{sha99} 
and \citet{gre01}. The interstellar X-ray absorption is taken into 
account by multiplying the {\tt tbabs} model \citep{wil00}. We adopt 
$N_{\rm H} = 7.4 \times 10^{21}$ cm$^{-2}$ \citep{mil06, tak08} 
as the column density of the Galactic absorption. The extinction of  
the OIR fluxes is also considered by combining the {\tt redden} 
model, in which the ratio of galactic extinctions 
$R_V = A_V/(A_V - A_B) = 3.1$ and the 
interstellar reddening $E(B-V) = 1.26$ are assumed. The value 
of $E(B-V)$ is derived from $N_{\rm H}$ and the relation 
$N_{\rm H}/E(B-V) = 5.8 \times 10^{21}$ cm$^{-2}$ mag$^{-1}$, 
determined by \citet{boh78}. 

The {\tt optxrplir} model assumes an inclination of $60^\circ$ 
in calculating the disk spectrum. To incorporate the inclination 
dependence of the disk emission, we set the normalization 
to $\cos i/\cos 60^\circ$ ($\approx 0.684$, in the case of 
GRO J1655$-$40 with $i = 70^\circ$). In addition, this 
model does not include relativistic smearing. Following 
\citet{don13}, we consider this effect by convolving our 
model with {\tt kerrconv} \citep{bre06}. The {\tt kerrconv} 
model uses $a_*$, $i$, inner and outer radii (in units of the 
ISCO radius: $R_{\rm ISCO}$) within which the relativistic 
effects are considered, the radial emissivity indices of inner 
and outer parts, and break radius (in units of $R_{\rm g}$) 
at which the emissivity index changes. We link $a_*$ in 
{\tt kerrconv} to that in {\tt optxrplir} and fix $i$ at $70^\circ$ 
and the emissivity indices at 3. The inner and outer radii 
are set to $1 R_{\rm ISCO}$ and $2 R_{\rm ISCO}$, 
respectively. In this way, we convolve the entire disk spectrum 
with a smearing kernel calculated from the radii between 
$1 R_{\rm ISCO}$ and $2 R_{\rm ISCO}$, 
in which a large fraction of the disk flux is produced\footnote{
The choice of the outer radius for the {\tt kerrconv} model 
does not substantially affect our results. In the case of a Schwarzschild 
black hole, $\gtrsim 10$\% of the total disk emission is produced between 
$1 R_{\rm ISCO}$ and $2 R_{\rm ISCO}$ \citep[][the percentage is slightly 
larger for $a_* \neq 0$]{nov73}. We confirm that the results of the spectral fits 
remain unchanged within the 90\% error ranges when we adopt 
5~$R_{\rm ISCO}$ and 10 $R_{\rm ISCO}$ (below which 
$\gtrsim 50$\% and $\gtrsim 70$\% of the flux are generated, 
respectively) as the outer radius.}, 
instead of integrating the smeared blackbody spectrum at each 
radius. As shown in \citet{don13}, the {\tt kerrconv*optx} 
model with this parameter setting gives a good approximation 
of relativistic disk models.

\subsubsection{SED fits for the normal high/soft state}
We first apply this model, 
"{\tt tbabs*redden(kerrconv*\\ optxrplir+bbodyrad)}", 
to the multi-wavelength SED on MJD 53446 in the normal 
high/soft state. Since X-ray spectra in that state typically 
have a power-law tail but no significant thermal 
Comptonization component, here we ignore the soft 
thermal Comptonization component in {\tt optxrplir} 
(i.e., $R_{\rm pl}$ is linked to $R_{\rm cor}$). 
As for irradiation, the strength of the reprocessed component 
is scaled with $F_{\rm out} = f_{\rm out} * (1-a_{\rm out})$ 
and it is difficult to constrain both $a_{\rm out}$ and 
$f_{\rm out}$ simultaneously in spectral fitting. Here, 
we assume $f_{\rm out} = 0.1$ and leave $a_{\rm out}$ 
as a free parameter to absorb uncertainties in the 
geometry of the outer disk.
The Eddington ratio ($L/L_{\rm Edd}$), BH spin $a_*$, and 
the photon index ($\Gamma$) and $R_{\rm pl}$ of the 
power-law tail are also allowed to vary. The cut-off 
energy of the power-law tail is set to be 1000 keV, 
a sufficiently high value compared with the energy 
range of the {\it RXTE} data. $R_{\rm out}$ is fixed at 
$10^{5.67} R_{\rm g}$ ($4.4 \times 10^{11}$~cm), which 
is the tidal radius of GRO J1655$-$40 given in 
\citet{sha98}\footnote{
This value corresponds to 80\% of the Roche lobe size. 
To test a different value, we leave $R_{\rm out}$ as a 
free parameter in the fit of the SED on MJD 53446, and 
use the best-fit value to fit the SED at the {\it Chandra} 
epoch. We confirm that this does not affect our conclusions.}. 
We extend the energy range to $10^{-4}$~keV--1000~keV 
in the following spectral fit, because {\tt kerrconv} is a 
convolution model.

The left panel of Figure~\ref{fig_fit} shows the results of fit  
on MJD 53446. The model well reproduces both the OIR fluxes 
and the X-ray spectrum, yielding $\chi^2/{\rm d.o.f.} = 99/105$. 
The free parameters are constrained to be $L/L_{\rm Edd} = 
0.1370^{+0.005}_{-0.006}$, $a_* = 0.76 \pm 0.03$, 
$R_{\rm pl} = 4.8 \pm 0.2 R_{\rm g}$, 
$\Gamma = 2.04 \pm 0.02$, and 
$a_{\rm out} = 0.974 \pm 0.03$. 
The strength of irradiated disk component is thus estimated 
to be $F_{\rm out} = (2.6 \pm 0.3) \times 10^{-3}$. 
We note that the high $a_{\rm out}$ value for $f_{\rm out} = 0.1$ 
is consistent with a highly ionized and reflective disk surface 
suggested by \citet{van81}, \citet{van83}, and \citet{gie09}, 
but combinations of somewhat smaller $a_{\rm out}$ and 
smaller $f_{\rm out}$ (where $f_{\rm out} > 2.3 \times 
10^{-3}$) may be also possible.
 
The best-fit model corrected for the interstellar 
absorption/extinction is plotted in Figure~\ref{fig_hssmodel}. 
It has a double-humped profile unlike the {\tt diskir} model, 
which instead produces a flat "shoulder" in the $\nu F_\nu$ form. 
This demonstrates the difference in radial dependence of 
the disk temperature. We find that the OIR fluxes are 
dominated by reprocessed emission from the outer edge of the 
disk irradiated by X-rays. The OIR bands are located at slightly 
lower frequencies than that of the peak in the reprocessed 
component, above and below which the relations between 
the incident X-ray luminosity ($L_{\rm X}$) and the 
luminosity of the irradiated outer disk ($L_{\nu, {\rm ir}}$) 
are given as 
$L_{\nu, {\rm ir}} \propto L_{\rm X}^{7/6}$ and 
$L_{\nu, {\rm ir}} \propto L_{\rm X}^{1/4}$, respectively  
\citep[see e.g.,][]{cor09}. 

\subsubsection{SED fits for the {\it Chandra} epoch 
in the hypersoft state}
Adopting the resultant parameters, we next apply the 
{\tt optxrplir} model to the SED on MJD 53461 in the 
hypersoft state. The right panel of Fig.~\ref{fig_fit} 
compares the SED on MJD 53461 with the same 
model on MJD 53446 in the high/soft state, except 
for the Eddington ratio, which is adjusted to fit the OIR 
fluxes. This gives a bolometric luminosity of 
$\approx$0.7 $L_{\rm Edd}$.
Since the OIR to X-ray flux ratio became lower 
on MJD 53461 than that in the normal high soft state 
(see Fig.~\ref{fig_lc}), the model does not fit the X-ray 
spectrum. 
We can clearly see that this model overestimates the 
soft X-ray flux by about 1 order of magnitude. 

The discrepancy seen in the right panel of Fig.~\ref{fig_fit} 
cannot be explained by absorption in the optically-thin wind 
detected by \citet{mil08}. We create a photoionized absorption 
model with XSTAR, employing the continuum spectral model 
in \citet{mil08} with a luminosity changed to 0.7 $L_{\rm Edd}$  
as an incident ionizing spectrum, and assuming the density of 
the wind as the value determined by them, 
$1 \times 10^{14}$~cm$^{-3}$. As the column density and 
ionization parameter, we adopt representative values in their 
photoionization models: $N_{\rm H} = 5 \times 10^{23}$~cm$^{-2}$ 
and $\xi = 10^4$~erg~cm~s$^{-1}$. However, we find this 
ionized absorption reduces only $\lesssim$3\% of the soft 
X-ray flux. Electron scattering in the absorber, which is 
not accounted for in the XSTAR code, can only decrease 
30\% of the radiation.
The discrepancy is neither canceled out by taking into account 
the decrease of the mass accretion rate in the inner disk region 
due to the mass loss in the wind. \citet{mil08} estimated 
the wind outflow rate to be $\sim$10\% of the mass accretion 
rate (assuming that the outflow velocity perpendicular to the 
disk plane is not much higher than the line-of-sight velocity), 
and in the more recent work by \citet{mil15}, it was estimated 
to be $\sim$30\% of the mass accretion rate.

One possibility is that the disk wind made irradiation 
stronger than the normal high/soft state. 
The difference in the X-ray versus OIR flux ratio is 
canceled out if $F_{\rm out}$ at the {\it Chandra} epoch 
is $\sim$4 times larger than that in the normal high/soft state.
It is difficult for disk winds to produce blackbody  
radiation by themselves, because they normally have 
much higher temperatures than those in the outer disks 
and are optically thin for free-free absorption. 
Instead, they can contribute to the irradiation of the 
disk by scattering X-ray photons down and enhancing the 
illumination of the outer disk \citep{gie09}. 
In this case, the geometry-dependent factor for 
irradiation can be expressed as 
\begin{equation}
f_{\rm out} \sim \frac{\Omega_{\rm wind}}{2\pi} 
(1- e^{-\tau_{\rm es}}),
\end{equation} 
where $\Omega_{\rm wind}$ 
and $\tau_{\rm es}$ represent the solid angle of the 
wind and the optical depth for scattering, respectively. 

In our case, however, it is unlikely that the OIR fluxes 
are enhanced by the optically-thin magnetic wind. 
The {\it Chandra} spectrum predominantly shows 
absorption features instead of prominent P-Cygni profiles, 
and hence $\Omega_{\rm wind}$ must be fairly small.
In addition, the disk at the location of the 
magnetic wind ($\sim 10^{9.7}$~cm) has a relatively high 
temperature and mainly produces photons with much higher 
energies than those in the OIR bands. The emitting region 
is too small to substantially contribute to the OIR fluxes. 
In the above modeling, the OIR fluxes are dominated by 
the reprocessed emission from the outer edge of the disk 
at $R_{\rm out} = 4.4 \times 10^{11}$~cm, where the 
temperature is $\approx 1$~eV. 
Using the radial dependence of the temperature in the 
irradiated disk, $T_{\rm irr} \propto r^{-3/7}$ 
\citep{cun76, vel13}, 
we estimate the disk temperature at the launching 
point to be $\approx 7$ times larger than that at 
$R_{\rm out}$. To investigate whether the reprocessed 
emission from the disk around the wind launching radius 
can explain the opposite behavior of the OIR and X-ray 
fluxes, we add a {\tt bbodyrad} component with a temperature 
of 7 eV to the model for MJD 53446, and compare it with 
the SED at the {\it Chandra} epoch. We find that the OIR 
flux levels cannot be reproduced unless we assume an 
unusually large solid angle at the launching point (more 
than 2 orders of magnitude larger than that of the disk). 

Moreover, it is difficult for an enhanced $F_{\rm out}$ 
alone to explain the trend in the soft X-ray spectrum:  
the inner disk temperature was not largely changed, 
while the observed X-ray flux was reduced. If the 
decline of the X-ray flux were produced simply by 
the decrease of the mass accretion rate in the inner 
disk region, the temperature should be reduced as 
well.
Considering all the above discussion, we conclude 
that an optically-thin wind is unlikely to be the main 
cause of the differences in SED between the normal 
high/soft state and the hypersoft state.

Instead, it seems more likely that
obscuring gas is present between the X-ray and optical 
emission regions, reducing the intrinsic X-ray spectrum. 
The gas is likely to be very highly ionized and affect the 
X-ray radiation by Compton-scattering, otherwise heavy 
absorption by less ionized material should modify the 
spectral shape in the soft X-ray band, inconsistent with 
the fact that it was kept almost constant in the hypersoft 
state (see Section~\ref{sec_longterm}). Thus, the origin 
of the gas could be a Compton-thick, equatorial disk wind, 
strongly scattering-out the X-ray photons from the line of sight. 

We then fit the multi-wavelength SED at the {\it Chandra} 
epoch using the soft thermal Comptonization component 
implemented in {\tt optxrplir}, to extract quantitative 
information of the observed very steep power-law tail 
and obtain a better SED model. We assume that the 
X-ray flux is scattered by an almost fully ionized, 
Compton-thick wind, while the OIR emission regions 
are located outside of the wind. Under the presence 
of such a wind, it is difficult to predict whether 
$F_{\rm out}$ becomes larger or smaller than that 
in the normal high/soft state, or remains constant. 
If the wind extends rather homogeneously over 
the outer disk region,  
the geometry-dependent factor $f_{\rm out}$ increases 
to $\sim \Omega_{\rm wind}(1- e^{-\tau_{\rm es}})/(2\pi)$ 
due to scattering in the wind. If the electron density is 
the highest at the wind launching point located at a much 
smaller radius than the outer edge, and if scattering 
predominantly occurs there, the wind reduces the 
X-ray photons illuminating the outer disk by 
$\sim \! e^{-\tau_{\rm es}}$.  
The optxrplir model uses the intrinsic X-ray luminosity 
to calculate the irradiated disk spectrum, and the decrease of 
illuminating photons is regarded as a decrease of $f_{\rm out}$. 
Here, we simply assume that $F_{\rm out}$ is not largely 
changed and use the same value in the high/soft state. 

We fixed the parameters, $a_*$ and $a_{\rm out}$ 
at the values determined in the spectral analysis 
of the normal high/soft state, and allowed the Eddington 
ratio and the parameters for the soft thermal Comptonization, 
$R_{\rm cor}$, $kT_{\rm e}$, and $\tau$ to vary. 
The power-law component is switched off by setting 
$R_{\rm pl}$ to less than the ISCO. To consider the 
observed ionized absorption in the wind, we utilize the 
photoionized absorption model created above, leaving  
the ionization parameter and column density as free 
parameters. 
The {\tt cabs} model is also incorporated to account 
for the effect that the dense wind (including both fully and 
partially ionized gas) scatters the X-ray photons out of 
the line of sight. We linked the column density of {\tt cabs} 
to that of the photoionized absorption model. The final fit 
model is {\tt tbabs*xabs*cabs*redden*(kerrconv*optxrplir + 
bbodyrad)}, where {\tt xabs} represents the photoionized 
absorption model. The {\tt xabs*cabs} and {\tt redden} 
components are turned off for the OIR and X-ray data, 
respectively, by fixing the column densities and reddening 
factor at negligibly small values. 

\begin{figure}
\plotone{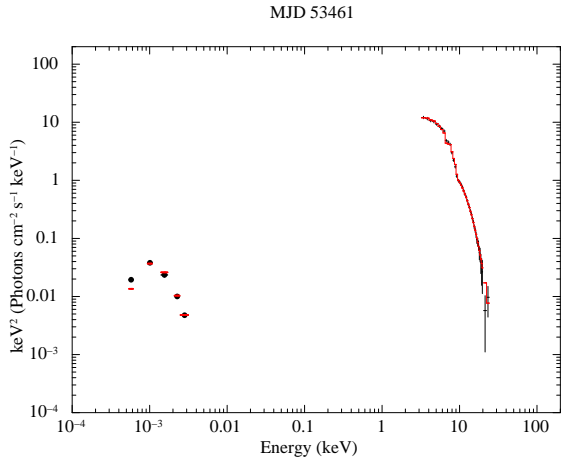}
\caption{Application of the {\tt optxrplir} model with the soft thermal Comptonization 
component to the SED on MJD 53461. The bottom panel shows the data vs. model 
ratio.
\label{fig_fit_chandra}}
\end{figure}

\begin{figure}
\plotone{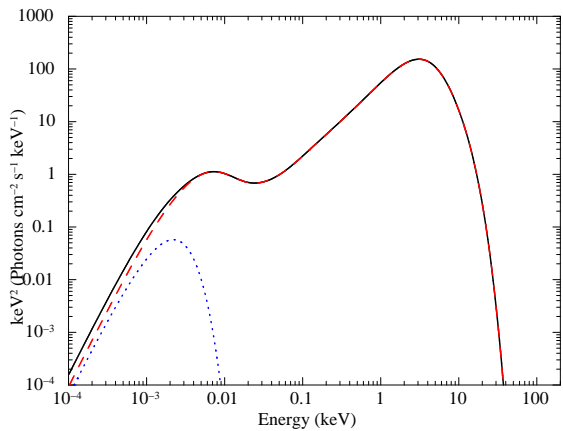}
\caption{The model presented in Figure~\ref{fig_fit_chandra}
corrected for neutral and ionized X-ray absorption, effects of 
Compton scattering produced by {\tt cabs}, and optical extinction. 
Dashed and dotted lines represent the {\tt kerrconv*optxrplir} 
and {\tt bbodyrad} components, respectively. 
\label{fig_model_chandra}}
\end{figure}

The condition that we try to model is quite complicated. 
We simply approximate the observed steep 
power-law component by the soft thermal Comptonization 
component included in the {\tt optxrplir} 
model. It assumes that the seed photons originate in 
the disk blackbody emitted from the inner disk region 
at radii within $R_{\rm cor}$ and that they are scatted 
by a one-zone cloud with given $\tau$ and $kT_{\rm e}$. 
However, the real location, geometry and internal structure 
of the Comptonized plasma are unknown. The origin of that 
component could be a dense corona or a hot flow located 
in the X-ray emitting region, or perhaps a Compton-thick 
disk wind in the outer disk region that strongly down-scatters 
X-ray photons and modifies the intrinsic spectral shape 
(see Section~\ref{sec_discussion}). 
Thermal Compton scattering in an optically-thick cloud 
with arbitrary structures has not been prepared as an 
XSPEC model, and it is difficult to fit the SED directly 
with realistic models. 
Moreover, we use one-zone homogeneous ionized 
absorption and scattering models to describe the disk 
wind, but the real structure would be more complex.
The main purpose of this analysis is not obtaining the 
precise values of the fit parameters, but investigating 
whether and how much optically-thick Compton scattering 
can reconcile the huge discrepancy shown in Figure~\ref{fig_fit}. 
Numerical calculation of radiative transfer employing 
various structures of the Comptonized cloud and disk wind 
would be needed for more detailed studies, which is beyond 
the scope of this article. We leave it as a future work. 

We find that the multi-wavelength SED can be 
characterized by the above model consisting of a disk 
with a luminosity of $L/L_{\rm Edd} \approx 0.74$ and 
a optically-thick, low-temperature Comptonization 
with $\tau \approx 5.3$ and $kT_{\rm e} \approx 1.4$~keV. 
Figure~\ref{fig_fit_chandra} plots the comparison of the 
data and the model, and Figure~\ref{fig_model_chandra} 
presents the same model corrected for neutral/photoionized 
absorptions and extinction. The column density of the 
{\tt cabs} and {\tt xabs} model is estimated to be 
$\approx 3.0 \times 10^{24}$ cm$^{-2}$ and $\xi$  
of the photoionized absorber is $\sim 10^4$ erg cm s$^{-1}$.
In this analysis, we have assumed that the outer disk 
emission is always seen in the OIR bands without obscuration. 
It could be partially hidden, however, if an optically-thick wind 
is present between the inner and outer disk regions. 
In this case, the intrinsic luminosity can be larger and 
the value estimated above can be regarded as a lower limit.
We find that, when the normalization of the {\tt kerrconv*optxrplir}  
component in the OIR bands is set to half of that in the X-ray 
band, the intrinsic disk luminosity goes up to $\sim 3 L_{\rm Edd}$. 
The model can still reproduce the observed X-ray spectrum 
by changing $R_{\rm cor}$ (which can be regarded as the 
strength of the soft thermal Comptonization component) from 
$26 R_{\rm g}$ to 50~$R_{\rm g}$, and the column density of 
the {\tt xabs} and {\tt cabs} components to 
$4.5 \times 10^{24}$~cm$^{-2}$.

\section{Discussion} \label{sec_discussion}

\subsection{The nature of the disk wind in GRO J1655$-$40}

\subsubsection{Summary of the unusual behaviors}
We have revisited the unusual wind detected with {\it Chandra} 
from the viewpoint of the long-term spectral and temporal 
evolutions in the X-ray and OIR bands. A number of peculiar 
behaviors were found to appear simultaneously in the 
hypersoft state around the {\it Chandra} observation, which 
are summarized as follows:
\begin{enumerate}

\item The observed OIR fluxes are monotonically increasing, 
whereas the X-ray flux starts decreasing before the 
{\it Chandra} observation. The B-band vs. X-ray flux ratio 
marks the minimal value just around the {\it Chandra} 
observation. 

\item The observed disk luminosity becomes lower, while 
the inner disk temperature remains almost the same. 
The standard $L_{\rm disk} \propto T_{\rm in}^4$ relation 
appears to break down \citep{don07}.

\item The power-law tail seen in the high/soft state spectra 
is replaced by a low-temperature optically-thick 
Comptonized component, while the OIR SED keeps 
almost the same shape.  

\item The absorption feature at $\approx$8 keV is always 
seen, and it becomes deeper as the observed X-ray flux 
decreases \citep{utt15}. 

\item Rapid variability is reduced as the observed X-ray flux 
decreased. The high-frequency break in the X-ray PDS 
moves downwards from $\approx 5$ Hz to $\approx 0.1$ Hz.  
The break frequency correlates tightly with the observed 
3--20 keV flux, while the energy dependence of the PSD 
shape is not strong \citep{utt15}.

\end{enumerate} 

These behaviors become most extreme at around the 
epoch of the wind detection with the {\it Chandra}/HETG. 
If they are not just a coincidence, they would be associated with 
the disk wind, since the ionized absorption feature always seen 
at $\approx 8$ keV suggests that the wind was steadily present 
in the hypersoft state. 

\subsubsection{Interpretation of the long-term trend in the X-ray and OIR bands}

As already discussed, the evolution of the OIR and X-ray fluxes 
cannot be explained by an increase of the mass accretion rate 
alone. In this case, the X-ray flux should have monotonically 
increased as well as the OIR fluxes. 
The decrease of the X-ray flux is too large to be canceled 
out by absorption and/or mass loss in the optically-thin disk 
wind detected with {\it Chandra}. It is also unlikely that an 
additional contribution from synchrotron emission from jets 
led to increase the OIR fluxes, since \citet{mig07} suggest 
that jets were quenched in the hypersoft state. 

One possible interpretation of the high OIR versus X-ray 
flux ratio in the hypersoft state is that reprocessed 
emission from the outer disk became stronger than 
the normal high/soft state. As discussed in \citet{gie09}, 
powerful disk winds can enhance the irradiation 
of the disk. The flux ratio at the {\it Chandra} epoch 
could be explained if $F_{\rm out}$, which is scaled with 
$f_{\rm out} * (1- a_{\rm out})$, was $\approx$4 times 
larger than that in the normal high/soft state. This means 
that, if $a_{\rm out}$ was constant, $f_{\rm out}$ is 0.4 in 
the hypersoft state. This value seems somewhat too large, 
given that no strong emission lines are seen in the 
{\it Chandra} spectrum. 
Instead, a factor of 4 increase in $F_{\rm out}$ may be 
achieved if $a_{\rm out}$ decreased somehow in the 
hypersoft state. It may be also possible that the actual 
$a_{\rm out}$ and $f_{\rm out}$ values in the normal 
high/soft state are smaller than what we estimated and 
that $f_{\rm out}$ increased in the hypersoft state.  
As explained in Section~3.2.2, we cannot constrain 
$f_{\rm out}$ and $a_{\rm out}$ simultaneously, and 
we simply fixed $f_{\rm out}$ at 0.1 in modeling the 
high/soft state spectrum. 
However, the launching radius of the wind estimated in 
\citet{mil08} is relatively small and thus its temperature is 
much higher than the outer edge of the disk, around which 
the OIR photons are emitted (see Section~3.2.3). 
It is difficult for the disk emission enhanced by that 
wind to contribute to the OIR fluxes,  
unless we consider an unusually large emitting region.

In addition, the larger-$F_{\rm out}$ interpretation 
alone cannot explain the peculiar behavior of the soft 
X-ray spectrum in the hypersoft state. If this were the 
case, the observed X-ray spectrum should directly reflect 
the intrinsic disk emission, and the trend in the disk 
luminosity versus inner disk temperature should indicate 
that the inner disk structure in the hypersoft state is different 
from that in the high/soft state, in which the inner edge of 
the standard disk is believed to extend to the ISCO stably 
\citep[e.g.,][]{ebi93, ste10, shi11a}. Some previous studies 
suggest that the disk is truncated at similar luminosities: 
in the bright low/hard state above $\sim 0.01 L_{\rm Edd}$ 
\citep[e.g., ][]{don07, shi11b, yam13, kol14, shi14, pla15} 
and in the very high state \citep{tam12, hor14} over 
$\sim 0.2 L_{\rm Edd}$ \citep{kub04}. 
However, the properties of the hypersoft state are nothing 
like these states; the X-ray spectra in the hypersoft state 
are much softer and the variability power on time scales of 
$\sim$0.01--10~s is more than one order of magnitude lower 
\citep[see e.g.,][for spectral and timing properties in each 
state]{mcc06,don07}. Moreover, disk truncation in these 
states produces a larger disk luminosity with a fixed inner 
temperature, which is the opposite behavior to that observed 
in the hypersoft state.

Instead, the evolution of the multi-wavelength SED could 
better described if $F_{\rm out}$ is not so largely 
changed between the high/soft and hypersoft 
states and obscuring gas is present between the 
inner and outer disk regions, intercepting the line 
of sight and significantly reducing the observed 
X-ray luminosity.
The obscuring material could be located in an equatorial disk 
wind. Strong Compton scattering in disk winds can reduce the 
intrinsic flux significantly, when we view them from a high 
inclination. Recent radiative transfer calculations have 
suggested that the scattered/reflected component of 
massive winds significantly contributes to the observed 
SEDs \citep{sim10a, sim10b}. Some observational hints 
of wind reprocessing have already been found in low-mass 
X-ray binaries \citep{dia12, dia14} implying that broad 
emission lines, blended with absorption features, could be 
explained by the reprocessed component of the winds. 
In our case, however, the Compton-thick part  
in the wind would be almost completely ionized. 
Such a wind can scatter out and reduce the intrinsic X-ray 
photons in the line of sight, without changing the spectral 
shape in the soft X-ray band nor generating additional 
significant atomic absorption and/or emission structures. 
As shown by our SED analysis in Section~\ref{sec_spec}, 
this scenario can well explain the unusual trend in the disk 
spectrum. Using the OIR fluxes as a tracer of the reprocessed 
luminosity in the outer disk, we have demonstrated that the 
apparent X-ray flux at the {\it Chandra} epoch can be 
$\gtrsim$1 order of magnitude lower than what is expected.

Similar very soft X-ray spectra with no or only a weak 
non-thermal power-law component are sometimes 
observed in other BH low-mass X-ray binaries, such as 
GRS 1915$+$105, XTE J1550$-$564, and GX 339$-$4 
\citep[e.g.,][]{kub04, zdz04}, in their brightest phase. 
They appear in the luminosity range from 
$\sim 0.4 L_{\rm Edd}$ to $\sim 1 L_{\rm Edd}$, which is 
consistent with the luminosity of GRO J1655$-$40 at the 
{\it Chandra} epoch that we estimated in the SED modeling. 
In previous works, the faint and/or steep power-law tail
above $\sim$10~keV was usually considered to be 
produced in the inner disk region, and the entire X-ray 
spectra in that phase were often reproduced by models 
composed of the standard disk and its Comptonized corona, 
or the so-called slim disk model, which describes accretion 
flows at very high mass accretion rates. 

In the case of GRO J1655$-$40, the scattered-in 
component of the Compton-thick disk wind could also 
contribute to the observed Comptonization component. 
The Compton temperature of the disk wind is estimated 
to be $T_{\rm C} \sim$1 keV, by assuming the unabsorbed 
{\tt optxrplir} model in Figure~\ref{fig_model_chandra} 
as the intrinsic spectrum. This is comparable to the 
electron temperature of the soft Comptonization 
component. The fairly large difference between the 
optical depths of the scattered-in component ($\tau \approx 5$, 
modeled by {\tt optxrplir}) and the scattered-out component 
($\tau \approx 2$, estimated from the column density of 
the {\tt cabs} model, $N_{\rm H} = 3 \times 10^{24}$~cm$^{-2}$), 
may suggest that different Comptonization components  
contribute to the soft component.
However, considering uncertainties in the spectral modeling, 
we do not rule out the possibility that the total flux of the soft 
component is explained by scattering in the disk wind alone.
We also note that the irradiated-disk emission enhanced by 
this Compton-thick wind would not contribute to the OIR fluxes 
greatly. This is because the wind is likely located at a much 
smaller radius than the outer edge of the disk (see the next section) 
and the relatively high disk temperature at that position requires 
a too large solid angle to produce the observed OIR emission 
(see also Section~3.2.3). 

\subsubsection{Where is the hidden Compton-thick wind located?}
\label{sec_windloc}
Considering the definition of the ionization parameter, 
we can constrain the position of the Compton-thick, almost 
completely ionized part in the disk wind, with respect to the 
less ionized part producing absorption lines. The radial 
dependence of the ionization parameter is derived from 
Equation~(\ref{eq_xiparam}) as $\xi (R) \propto 1/(n R^2)$. 
If the density profile can be written as  
$n \propto R^{-\alpha}$, a wind is more ionized 
at larger radii under the condition $\alpha > 2$, 
while it has the opposite dependence when $\alpha < 2$. 
The former condition can be established 
if the disk wind is accelerated in its outer parts. The mass 
loss rate $\dot{M}_{\rm wind}$ is described as 
\begin{equation}
\dot{M}_{\rm wind} \propto R^2 n v_{\rm wind} \Omega,
\end{equation}
where $\Omega$ and $v_{\rm wind}$ represent the solid 
angle and velocity of the wind, respectively. 
A positive radial dependence of the outflow velocity 
corresponds to $\alpha > 2$, when the mass loss rate 
and the solid angle are kept constant. Oppositely, the 
latter condition $\alpha < 2$ can be realized when the 
wind is decelerated. 

The former case means that the Compton-thick region was 
located at a larger radius than that producing absorption 
lines, and that the latter part directly percepts the intrinsic 
X-ray luminosity. 
The luminosity that we estimated is about 20 times larger than 
the ionizing X-ray luminosity assumed in the photoionized 
plasma modeling in previous works. Since the launching 
radius depends on the square root of the ionizing luminosity 
(see Equation~(\ref{eq_xiparam})), it can be larger than that 
estimated in \citet{mil08} by a factor of $\gtrsim$ 4.4. 
In addition, radiation pressure reduces the local effective 
gravitational force by $(1- L/L_{\rm Edd}) \lesssim 0.3$, 
and thereby the Compton radius is decreased by the same 
factor. Thus, the two factors can compensate the 
1-order-of-magnitude difference between the wind launching 
radius estimated from the {\it Chandra}/HETG data and the 
minimum radius at which a thermal wind can be driven. 
The Compton temperature estimated from our SED model 
($T_{\rm C} \sim 1$ keV) gives the critical luminosity 
$L_{\rm crit} = 0.03 L_{\rm Edd}$, and hence the intrinsic 
luminosity is sufficiently large to launch a thermal wind. 

By contrast, the X-ray photons illuminating the less 
ionized part of the wind should be already reduced 
in the latter case, $\alpha < 2$. We suggest this condition 
would be more plausible than $\alpha > 2$, considering 
the column density of the scattered-out component 
($N_{\rm H} = 3 \times 10^{24}$ cm$^{-2}$). Approximating 
the density as $n \sim N_{\rm H} \Delta R$, where $\Delta R$ 
is the size of the Compton-thick part in the line-of-sight 
direction, and assuming the ionized parameter of that 
part as $\xi \gtrsim 10^6$ erg cm s$^{-1}$, we derive 
\begin{equation}
R \sim \frac{L_{\rm X}}{N_{\rm H} \xi} \frac{\Delta R}{R} \lesssim 3 \times 10^8 \left(\frac{\Delta R}{R} \right) 
\left(\frac{L_{\rm X}}{1 L_{\rm Edd}} \right) {\rm cm}.
\end{equation}
This value is smaller than the wind launching radius 
estimated by \citet{mil08}. A similar trend in $\xi$ has 
recently been found in the re-analysis of the 
{\it Chandra} data by \citet{mil15}. From a detailed 
analysis of the absorption line profile, they detected 
multiple components in the wind, where a more 
ionized part was located at a smaller radius. 

After \citet{mil06} first reported the {\it Chandra}/HETG result, 
the unusual wind in GRO J1655$-$40 has been revisited 
many times from both the observational and theoretical points 
of view \citep{net06, mil08, kal09, luk10, nei12, hig15}. 
Many previous studies 
were based on the assumption that the wind was optically thin 
and launched at a few percent of the Eddington luminosity, 
in constructing photoionized plasma models and performing 
hydrodynamic simulations. If the wind was Compton-thick 
and formed at near- or super-Eddington luminosity, 
the effects of scattering in the wind and radiation pressure 
should be considered to discuss the wind driving mechanism. 
As described above, the launching radius of a thermal 
driven wind is smaller at a higher luminosity because radiation 
pressure reduces the effective gravity. A radiation pressure 
driven wind can be launched at even smaller radii above 
Eddington luminosity. Further numerical simulations 
and radiation transfer calculations at high mass accretion 
rates including the effects of Compton scattering 
would be needed to test whether a Compton-thick, 
almost completely ionized wind can be really launched 
and reproduce the observed properties.

\subsubsection{X-ray short-term timing properties} 
Similar hypersoft X-ray spectra are sometimes detected 
in the high-mass X-ray binary Cyg X-3 as well 
(\citealt{kol10}; see \citealt{utt15} for comparison 
of the two sources). This system always exhibits 
peculiar spectral states and lacks rapid variability 
on shorter time scales than $\sim$0.1~Hz. 
\citet{zdz10} succeeded in describing spectral shapes 
and the timing properties in Cyg X-3 by considering 
Compton-thick scattering plasma, likely formed by 
dense stellar winds from its Wolf-Rayet companion star, 
with a low temperature (3~keV) and a large optical depth 
($\tau \sim 7$) like those of the observed thermal 
Comptonization component in GRO J1655$-$40. 
They found that the break frequency ($\nu_{\rm b}$) 
in the X-ray PDS is scaled by the size ($R_{\rm scat}$) 
and the optical depth of the scattering cloud that reduces 
the intrinsic fast variability, through the following relation: 
\begin{equation}
R_{\rm scat} \simeq \frac{c}{2 \tau \nu_{\rm b}}.
\end{equation}

This could be applied to GRO J1655$-$40 in the 
hypersoft state, if its low break frequency is caused 
by scattering in the Compton-thick disk wind. 
Using the optical depth for scattering in the line of sight 
($\tau \sim 2$) and the break frequency (0.1~Hz) at the 
{\it Chandra} epoch, we derive the size of the scattering 
gas to be $R \sim 10^5 R_{\rm g} \approx 10^{11}$~cm, 
corresponding to $\sim 0.2 R_{\rm C}$.
This value is much larger than the radius of the Compton-thick 
wind estimated in Section~\ref{sec_windloc} and appears to 
favor a different interpretation that the Compton-thick gas 
extends to the outermost disk region.  
We note, however, that there are other possible explanations 
for the suppression of the short-term X-ray variations in the 
hypersoft state. As discussed in \citet{utt15}, the PDS profile 
may be generated by the intrinsic variation of the accretion 
flow, or variable scattering in the wind due to density 
inhomogeneities (see Section~\ref{sec_ULX}). 

\subsection{A previously detected wind in GRO J1655$-$40}
A similar or more extreme disk wind might be observed 
in a previous outburst as well. The first detection of ionized 
absorption lines in GRO J1655$-$40 was brought 
by {\it Advanced Satellite for Cosmology and Astrophysics} 
({\it ASCA}; \citealt{tan94}) in the outburst starting from 1994 
\citep{ued98}. Ionized Fe-K absorption lines were detected 
in two epochs, when the source exhibited similar X-ray 
continuum spectra but had different observed X-ray fluxes 
by about one order of magnitude. The ionized absorber 
was more ionized when the source was brighter, as was 
seen in the 2005 outburst \citep{dia07, nei12}. 
The observed X-ray flux in the {\it Chandra} observation falls 
between those at the two epochs reported by \citet{ued98}. 

A sudden flux drop was detected in the {\it ASCA} 
observation at the lower flux (in 1994 August). The duration 
of the event was several hours, which was much longer 
than those observed in different periods 
\citep[on time scales of second or minute][]{kuu98, sha07}.
Interestingly, in this "dip" event, the source kept almost 
the same continuum shape as outside it, unlike normal 
absorption dips in low-mass X-ray binaries, which 
accompany spectral hardening 
\citep[e.g.,][]{kuu98, hom05, dia06, kuu13, shi13}. 
In the dipping period, the ionized Fe-K absorption lines 
disappeared, while a deep absorption edge appeared at 
$\approx$ 7 keV. This spectral behavior, which was described 
with double partial-covering absorbers in \citet{tan03}, is 
reminiscent of the behavior during the hypersoft state in 
the 2005 outburst. The similarity might suggest that the 
unusual "dip" event was produced by absorption 
and scattering in a more extreme wind with an even 
larger column density and an lower ionization parameter 
than that observed in 2005, and that a large fraction of 
the intrinsic X-ray photons was scattered out in the wind. 
There is, however, no information of the spectral 
shape in the hard X-ray band above 10 keV 
for the {\it ASCA} data, hampering further comparison. 

\subsection{Implications for disk winds in other luminous X-ray binaries}
\label{sec_ULX}

If these peculiar behaviors in GRO J1655$-$40 seen around 
the {\it Chandra} observation are indeed attributed to an 
optically-thick disk wind launched at a luminosity close to 
the Eddington limit, they would carry important implications 
to massive winds thought to be formed at extreme mass 
accretion rates. Recent radiation (magneto-)hydrodynamic 
simulations predict radiation-pressure dominated, optically-thick 
disk winds to be driven in supercritical accretion disks 
\citep[e.g.,][]{ohs05, ohs11, dot11}. Although the ionized 
Fe-K absorption lines like those observed in Galactic BH 
binaries have never been observed in ultraluminous X-ray 
sources \citep[ULXs;][]{wal12, wal13}, \citet{mid14} argued 
that uneven spectral structures often seen at $\approx$1 keV 
could be explained by broad blueshifted absorption lines. 
Recent deep optical spectroscopy revealed that the broad 
emission lines of H and He observed in ULXs (and also 
SS 433, which is the only known Galactic X-ray binary 
thought to power persistent supercritical accretion; see 
\citealt{fab04} and references therein) likely originate in 
dense, strong winds launched on supercritical accretion disks 
\citep{fab15}\footnote{We note that these optical emission 
lines were emitted from much less ionized gas, presumably 
located at more distant parts in the winds than the optically-thick 
wind in GRO J1655$-$40 that we have considered.}.  
Moreover, \citet{gla09} suggest that massive optically-thick 
winds could be the origin of the soft Comptonization component 
usually detected in ULXs. It has a low temperature ($\approx$3~keV) 
and a large optical depth ($\tau \approx 5$--$30$), similar to 
those estimated for the thermal Comptonization component 
in GRO J1655$-$40. Thus, the unusual GRO J1655$-$40 wind, 
formed at around the Eddington luminosity, may be smoothly 
connected to massive outflows from supercritical accretion 
flows. 
More recently, \citet{sor15} and \citet{urq15} have proposed 
a unified picture of the ULXs and the ultraluminous supersoft X-ray 
sources (ULSs, whose spectra are dominated by a blackbody-like 
component with a temperature of $< 0.1$~keV and have no hard tail), 
in terms of the effective optical depth ($\tau_{\rm eff}$) of the 
supercritical winds: in ULSs, the winds are effectively optically thick 
and we see only a blackbody emission, while ULX winds are  
optically thick to Compton scattering but $\tau_{\rm eff} \lesssim 1$, 
and thus produce a Comptonized component in the X-ray band. 
We note that $\tau_{\rm eff} << 1$ in the case of GRO J1655$-$40, 
when a temperature of $\sim$1 keV and a density of 
$\sim 10^{16}$~cm$^{-3}$ are assumed to estimate the opacity 
of free-free absorption in the wind.

Recent simulations by \citet{tak13} demonstrated that the 
winds driven by supercritical accretion have a clumpy structure, 
which can generate variability. Spectral/flux variation in ULXs that 
could be explained by the clumpy disk winds has been observed 
on various wavelengths and time scales, such as X-ray flux 
on a time scale of several tens of second \citep{mid11} 
and optical emission lines on $\lesssim$1-day time scales 
\citep[][]{fab15}. 
As pointed out by \citet{utt15}, the clumpiness of the wind 
may contribute to the observed evolution of high-frequency 
variability in GRO J1655$-$40. If the disk wind have the 
Keplerian velocity in the rotating direction, 0.1-s and 10-s 
variations correspond to the blob sizes of $10 R_{\rm g}$ 
and $10^2  R_{\rm g}$ located at $10^4  R_{\rm g}$, 
respectively.
If the bend in the X-ray PDSs is produced by the absorption 
and scattering in a clumpy wind, variability at a lower 
frequency is generated by blobs with a larger size and/or 
located at a larger radius on the disk. In this case, the wind 
in the {\it Chandra} observation should have the largest 
blobs or be located at the most outer region during 
the hypersoft state, although it is not clear what are the 
mechanisms that made change the geometry and/or 
inner structures of the wind.

Wind absorption and scattering may also significantly affect 
the spectra of normal Galactic BH X-ray binaries, when they get close to 
the Eddington luminosity. At high luminosities, 
the dependence of the disk luminosity ($L_{\rm disk}$) on the 
inner disk temperature ($T_{\rm in}$) in the high/soft state 
start departing from the $L_{\rm disk} \propto T_{\rm in}^4$ 
track for the untruncated standard disk \citep[see][]{don07}. 
High inclination sources with absorption dips and winds,   
including GRO J1655$-$40 and 4U 1630$-$47, deviate 
downward from the relation in the $L_{\rm disk}$ vs. 
$T_{\rm in}^4$ plane with increasing luminosity, while low 
inclination systems like GX 339$-$4 and LMC X-3 do not 
show such a bend but sometimes go slightly upward. 
This apparent inclination-dependent discrepancy might be 
associated with optically-thick disk winds and explained 
as follows: 
in high inclination sources, we see the inner disk region through 
the disk wind, and the intrinsic X-ray emission is reduced due 
to scattering out of the line of sight. In contrast, lower inclination 
sources have larger contribution of the components scattered into 
the line of sight, which enhance the X-ray emission from the 
inner disk region. 
Moreover, Compton down-scattering in the wind could decrease 
the observed inner disk temperature. If this is the case, the winds 
should be almost fully ionized, otherwise significant emission 
and absorption lines could be seen in low and high inclination 
systems, respectively. The intrinsic luminosity  
at the {\it Chandra} epoch that we estimated is consistent with 
the energy range that the deviation is seen in other systems, 
and the GRO J1655$-$40 wind may be linked to that behavior 
as an extreme case. 

\section{Conclusion}
In this article, we proposed the possibility that the unusual 
wind in GRO J1655$-$40 was Compton-thick and almost 
completely ionized. We demonstrated that the 
multi-wavelength SED can be reproduced by considering 
the effects of scattering in the wind. We found that the 
intrinsic luminosity could be close to or above Eddington, 
which is much higher than that inferred 
from the observed X-ray flux. If this is the case, Compton 
heating and radiation pressure could be more important 
to drive the wind. To test this possibility, 
detailed radiation transfer calculation should be made by 
fully taking into account the effects of scattering in the wind.

We note a paper by \citet{nei16}, which has appeared 
very recently. They analyzed X-ray and OIR data of GRO J1655-40 
obtained in the similar period and drew a similar conclusion to 
ours that the wind was Compton thick. 
Their interpretation is that the wind was much more powerful  
and launched at much more extreme mass accretion rate ($\sim 30$ 
times as high as the Eddington rate) than that in our interpretation,  
and that most of the observed OIR fluxes in the hypersoft state 
were produced by the emission from the cool photosphere of 
the wind itself, instead of the emission from irradiated disk. 
As discussed in Section~3.2.3, the wind in our model is likely 
to be optically thin for free-free absorption in the OIR bands, 
but such a dense wind as suggested by \citet{nei16} 
may be effectively thick in the OIR bands and generate 
blackbody emission.

\acknowledgments

We are grateful to Michelle Buxton and Charles Bailyn 
for giving us the optical and near-infrared data obtained 
with the SMARTS. We thank Roberto Soria, Kazuo Makishima, 
and Joey Neilsen for fruitful discussions, and the anonymous 
referee for useful comments that helped to improve the paper. 
MS acknowledges support by the Special Postdoctoral 
Researchers Program at RIKEN. CD acknowledges STFC 
support from grant ST/L00075X/1. This work is partly 
supported by a Grant-in-Aid for JSPS Fellows for Young 
Researchers (MS) and for Scientific Research 26400228 (YU). 
This research has made use of data obtained from the High 
Energy Astrophysics Science Archive Research Center (HEASARC), 
provided by NASA's Goddard Space Flight Center.


\begin{thebibliography}{}

\bibitem[Begelman et al.(1983)]{beg83} Begelman, M.~C., McKee, C.~F., \& Shields, G.~A. \ 1983, \apj, 271, 70
\bibitem[Bohlin et al.(1978)]{boh78} Bohlin, R.~C., Savage, B.~D., \& Drake, J.~F.\ 1978, \apj, 224, 132
\bibitem[Boirin et al.(2005)]{boi05} Boirin, L., M{\'e}ndez, M., D{\'{\i}}az Trigo, M., Parmar, A.~N., \& 
Kaastra, J.~S.\ 2005, \aap, 436, 195
\bibitem[Boirin \& Parmar(2003)]{boi03} Boirin, L., \& Parmer, A.~N.\ 2003, \aap, 407, 1079
\bibitem[Brenneman \& Reynolds(2006)]{bre06} Brenneman, L.~W., \& Reynolds, C.~S.\ 2006, \apj, 652, 1028
\bibitem[Buxton \& Bailyn(2004)]{bux04} Buxton, M., \& Bailyn, C.\ 2004, \apj, 615, 880 
\bibitem[Buxton et al.(2005)]{bux05} Buxton, M., Bailyn, C., \& Maitra, D.\ 2005, ATel, \#418, 1
\bibitem[Corbel \& Fender(2002)]{cor02} Corbel, S., \& Fender, R.~P.\ 2002, \apj, 573, L35
\bibitem[Coriat et al.(2009)]{cor09} Coriat, M., Corbel, S., Buxton, M.~M., et al.\ 2009, \mnras, 400, 123
\bibitem[Cunningham(1976)]{cun76} Cunningham, C.\ 1976, \apj, 208, 534
\bibitem[D{\'{\i}}az Trigo et al.(2006)]{dia06} D{\'{\i}}az Trigo, M., Parmar, J., Boirin, L., M{\'e}ndez, M., 
\& Kaastra, J.~S.\ 2006, \aap, 445, 179
\bibitem[D{\'{\i}}az Trigo et al.(2007)]{dia07} D{\'{\i}}az Trigo, M., Parmar, A. N., Miller, J., Kuulkers, E., \& 
Caballero-Garc{\'{\i}}a, M.~D.\ 2007, \aap, 462, 657
\bibitem[D{\'{\i}}az Trigo et al.(2009)]{dia09} D{\'{\i}}az Trigo, M., Parmar, A.~N., Boirin, L., et al.\ 2009, \aap, 493, 145
\bibitem[D{\'{\i}}az Trigo et al.(2012)]{dia12} D{\'{\i}}az Trigo, M., Sidoli, L., Boirin, L., \& Parmar, A.~N.\ 2012, \aap, 543, 50
\bibitem[D{\'{\i}}az Trigo \& Boirin(2013)]{dia13} D{\'{\i}}az Trigo, M., \&  Boirin, L.\ 2013, Acta Polytechnica, 53, 659
\bibitem[D{\'{\i}}az Trigo et al.(2014)]{dia14} D{\'{\i}}az Trigo, M., Migliari, S., Miller-Jones, J.~C.~A., \& 
Guainazzi, M.\ 2014, \aap, 571, 76
\bibitem[Done et al.(2007)]{don07} Done, C., Gierli\'nski, M., \& Kubota, A.\ 2007, \aapr, 15, 1
\bibitem[Done et al.(2012)]{don12} Done, C., Davis, S.~W., Jin, C., Blaes, O., \& Ward, M.\ 2012, 
\mnras, 420, 1848
\bibitem[Done et al.(2013)]{don13} Done, C., Jin, C., Middleton, M., \& Ward, M.\ 2013, \mnras, 434, 1955
\bibitem[Dotan \& Shaviv(2011)]{dot11} Dotan, C., \& Shaviv, N.~J.\ 2011, \mnras, 413, 1623
\bibitem[Ebisawa et al.(1993)]{ebi93} Ebisawa, K., Makino, F., Mitsuda, K., et al.\ 1993, \apj, 403, 684
\bibitem[Fabrika et al.(2015)]{fab15} Fabrika, S., Ueda, Y., Vinokurov, A., Sholukhova, O., \& Shidatsu, M.\ 2015, 
Nature Physics, 11, 551
\bibitem[Fabrika(2004)]{fab04} Fabrika, S.\ 2004, Astrophysics and Space Physics Reviews, 12, 1
\bibitem[Frontera et al.(2001)]{fro01} Frontera, F., Palazzi, E., Zdziarski, A.~A., et al.\ 2001, \apj, 546, 1027
\bibitem[Gandhi et al.(2010)]{gan10} Gandhi, P., Dhillon, V.~S., Durant, M., et al.\ 2010, \mnras, 407, 2166
\bibitem[Gierli{\'n}ski et al.(2008)]{gie08} Gierli{\'n}ski, M., Done, C., \& Page, K.\ 2008, \mnras, 388, 753
\bibitem[Gierli{\'n}ski et al.(2009)]{gie09} Gierli{\'n}ski, M., Done, C., \& Page, K.\ 2009, \mnras, 392, 1106
\bibitem[Gladstone et al.(2009)]{gla09} Gladstone, J.~C., Roberts, T.~P., \& Done, C.\ 2009, \mnras, 397, 1836
\bibitem[Greene et al.(2001)]{gre01} Greene, J., Bailyn, C.~D., \& Orosz, J.~A.\ 2001, \apj, 554, 1290
\bibitem[Higginbottom \& Proga(2015)]{hig15} Higginbottom, N., \& Proga, D.\ 2015, \apj, 807, 107
\bibitem[Hjellming \& Rupen(1995)]{hje95} Hjellming, R.~M., \& Rupen, M.~P.\ 1995, \nat, 375, 464
\bibitem[Homan et al.(2005)]{hom05} Homan, J., Miller, J.~M., Wijnands, R., et al.\ 2005, \apj, 623, 383
\bibitem[Hori et al.(2014)]{hor14} Hori, T., Ueda, Y., Shidatsu, M., et al.\ 2014, \apj, 790, 20
\bibitem[Hynes et al.(2003)]{hyn03} Hynes, R.~I., Haswell, C.~A., Cui, W., et al.\ 2003, \mnras, 345, 292
\bibitem[Ibragimov et al.(2005)]{ibr05} Ibragimov, A., Poutanen, J., Gilfanov, M., Zdziarski, A.~A. \& Shrader, 
C.~R.\ 2005 \mnras, 362, 1435
\bibitem[Kallman et al.(2009)]{kal09} Kallman, T.~R., Bautista, M.~A., Goriely, S., et al.\ 2009, \apj, 701, 865
\bibitem[Kolehmainen et al.(2014)]{kol14} Kolehmainen, M., Done, C., \& D{\'{\i}}az Trigo, M.\ 2014, \mnras, 437, 316
\bibitem[Koljonen et al.(2010)]{kol10} Koljonen, K.~I.~I., Hannikainen, D.~C., McCollough, M.~L., Pooley, G.~G., 
\& Trushkin, S.~A.\ 2010, \mnras, 406, 307
\bibitem[Kotani et al.(2000)]{kot00} Kotani, T., Ebisawa, K., Dotani T., et al.\ 2000, \apj, 539, 413
\bibitem[Kubota et al.(2007)]{kub07} Kubota, A., Dotani, T., Cottam, J., et al.\ 2007 \pasj, 59, 185
\bibitem[Kubota \& Makishima(2004)]{kub04} Kubota, A., \& Makishima, K.\ 2004, \apj, 601, 428
\bibitem[Kuulkers et al.(1998)]{kuu98} Kuulkers, E., Wijnands, R., Belloni, T., et al.\ 1998, \apj, 494, 753 
\bibitem[Kuulkers et al.(2013)]{kuu13} Kuulkers, E., Kouveliotou, C., Belloni, T., et al.\ 2013, \aap, 552, 32 
\bibitem[Luketic et al.(2010)]{luk10} Luketic, S., Proga, D., Kallman, T.~R., Raymond, J.~C., 
\& Miller, J.~M.\ 2010, \apj, 719, 515
\bibitem[McClintock \& Remillard(2006)]{mcc06} McClintock, J.~E., \& Remillard, R.~A.\ 2006, in Compact Stellar X-Ray Sources,
ed.\ W.~H.~G., Lewin \& M. van der Klis (Cambridge: Cambridge Univ. Press), 157
\bibitem[Middleton et al.(2011)]{mid11} Middleton, M.~J., Roberts, T.~P., Done, C., \& Jackson, F.~E.\ 2011, 
\mnras, 411, 644
\bibitem[Middleton et al.(2014)]{mid14} Middleton, M.~J., Walton, D.~J., Roberts, T.~P., \& Heil, L.\ 2014, \mnras, 438, L51
\bibitem[Migliari et al.(2007)]{mig07} Migliari, S., Tomsick, J.~A., Markoff, S.\ et al.\ 2007, \apj, 670, 610
\bibitem[Miller et al.(2006)]{mil06} Miller, J.~M., Raymond, J., Fabian, A.~C., et al.\ 2006, \nat, 441, 953
\bibitem[Miller et al.(2008)]{mil08} Miller, J. M., Raymond, J., Reynolds, C.~S., et al.\ 2008, \apj, 680, 1359
\bibitem[Miller et al.(2015)]{mil15} Miller, J.~M., Fabian, A.~C., Kaastra, J.~S., et al.\ 2015, \apj, 814, 87
\bibitem[Mitsuda et al.(1984)]{mit84} Mitsuda, K., Inoue, H., Koyama, K., et al.\ 1984, \pasj, 36, 741
\bibitem[Nakahira et al.(2012)]{nak12} Nakahira, S., Koyama, S., \& Ueda, Y., et al.\ 2012, \pasj, 64, 13 
\bibitem[Neilsen \& Lee(2009)]{nei09} Neilsen, J., \& Lee, J.~C.\ 2009, \nat, 458, 481
\bibitem[Neilsen et al.(2011)]{nei11} Neilsen, J., Remillard, R.~A., \& Lee, J.~C.\ 2011, \apj, 737, 69
\bibitem[Neilsen \& Homan(2012)]{nei12} Neilsen, J., \& Homan, J.\ 2012, \apj, 750, 27
\bibitem[Neilsen et al.(2016)]{nei16} Neilsen, J., Rahoui, F., Homan, J., \& Buxton, M.\ 2016, 
ArXiv:1603.04070
\bibitem[Netzer(2006)]{net06} Netzer, H.\ 2006, \apj, 652, 117 
\bibitem[Novikov \& Thorne(1973)]{nov73} Novikov, I.~D., \& Thorne K.~S.\ 1973, in 
Black Holes, ed.\  C., Dewitt, \& B.~S., Dewitt, (New York: Gordon \& Breach), 343
\bibitem[Ohsuga et al.(2005)]{ohs05} Ohsuga, K., Mori, M., Nakamoto, T., \& Mineshige, S.\ 2005, \apj, 628, 368
\bibitem[Ohsuga \& Mineshige(2011)]{ohs11} Ohsuga, K., \& Mineshige, S.\ 2011, \apj, 736, 2
\bibitem[Orosz \& Bailyn(1997)]{oro97} Orosz, J.~A., \& Bailyn, C.~D.\ 1997, \apj, 477, 876
\bibitem[Plant et al.(2015)]{pla15} Plant, D.~S., Fender, R.~P., Ponti, G., Mu{\~n}oz-Darias, T., Coriat, M.\ 
2015, \aap, 573, 120
\bibitem[Ponti et al.(2012)]{pon12} Ponti, G., Fender, R.~P., Begelman, M.~C., et al.\ 2012, \mnras, 422, L11
\bibitem[Poutanen \& Veledina(2014)]{pou14a} Poutanen, J., \& Veledina, A.\ 2014, \ssr, 183, 61
\bibitem[Poutanen et al.(2014)]{pou14b} Poutanen, J., Veledina, A., \& Revnivtsev, M.~G.\ 2014, \mnras, 445, 3987
\bibitem[Russell et al.(2007)]{rus07} Russell, D.~M., Fender, R.~P., \& Jonker, P.~G.\ 2007, \mnras, 379, 1108
\bibitem[Shahbaz et al.(1998)]{sha98} Shahbaz, T., Charles, P.~A., \& King, A.~R.\ 1998, \mnras, 301, 382
\bibitem[Shahbaz et al.(1999)]{sha99} Shahbaz, T., van der Hooft, F., Casares, J., Charles, P.~A., \& 
van Paradijs, J.\ 1999 \mnras, 306, 89
\bibitem[Shaposhnikov et al.(2007)]{sha07} Shaposhnikov, N., Swank, J., Shrader, C.~R., et al.\ 2007, \apj, 655, 434
\bibitem[Shidatsu et al.(2011a)]{shi11a} Shidatsu, M., Ueda, Y., Nakahira, S., et al.\ 2011a, \pasj, 63, 803
\bibitem[Shidatsu et al.(2011b)]{shi11b} Shidatsu, M., Ueda, Y., Tazaki, F., et al.\ 2011b, \pasj, 63, 785
\bibitem[Shidatsu et al.(2013)]{shi13} Shidatsu, M., Ueda, Y., Nakahira, S., et al.\ 2013, \apj, 779, 26
\bibitem[Shidatsu et al.(2014)]{shi14} Shidatsu, M., Ueda, Y., Yamada, S., et al.\ 2014, \apj, 789, 100
\bibitem[Sim et al.(2010a)]{sim10a} Sim, S.~A., Miller, L., Long, K.~S., Turner, T.~J., \& Reeves, J.~N.\ 2010, 
\mnras, 404, 1369
\bibitem[Sim et al.(2010b)]{sim10b} Sim, S.~A., Proga, D., Miller, L., Long, K.~S., \& Turner, T.~J.\ 2010, \mnras, 
408, 1396
\bibitem[Soria \& Kong(2015)]{sor15} Soria, R., \& Kong, A.~K.~H.\ 2015, \mnras, in press
\bibitem[Steiner et al.(2010)]{ste10} Steiner, J.~F., McClintock, J.~E., Remillard, R.~A., et al.\ 2010, \apj, 718, L117
\bibitem[Sutton et al.(2014)]{sut14} Sutton, A.~D., Done, C., \& Roberts, T.~P.\ 2014, \mnras, 
444, 2415
\bibitem[Takahashi et al.(2008)]{tak08} Takahashi, H., Fukazawa, Y., Mizuno, T., et al.\ 2008, \pasj, 60, 69
\bibitem[Takeuchi et al.(2013)]{tak13} Takeuchi, S., Ohsuga, K., \& Mineshige, S.\ 2013, \pasj, 65, 88
\bibitem[Tamura et al.(2012)]{tam12} Tamura., M., Kubota, A., Yamada, S., et al.\ 2012, \apj, 753, 65
\bibitem[Tanaka et al.(1994)]{tan94} Tanaka, Y., Inoue, H., \& Holt, S.~S.\ 1994, \pasj, 46, L37
\bibitem[Tanaka et al.(2003)]{tan03} Tanaka, Y., Ueda, Y., \& Boller, T.\ 2003, \mnras, 338, L1
\bibitem[Ueda et al.(1998)]{ued98} Ueda, Y., Inoue, H., Tanaka, Y., et al.\ 1998, \apj, 492, 782
\bibitem[Ueda et al.(2001)]{ued01} Ueda, Y., Asai, K., Yamaoka, K., Dotani, T., \& Inoue, H.\ 2001 \apj, 556, L87
\bibitem[Ueda et al.(2009)]{ued09} Ueda, Y., Yamaoka, K., \& Remillard, R.\ 2009 \apj, 695, 888
\bibitem[Ueda et al.(2010)]{ued10} Ueda, Y., Honda, K., Takahashi, H., et al.\ 2010, \apj, 713, 257
\bibitem[Urquhart \& Soria (2015)]{urq15} Urquhart, R., \& Soria, R.\ 2015, \mnras, in press
\bibitem[Uttley \& Klein-Wolt(2015)]{utt15} Uttley, P., \& Klein-Wolt, M.\ 2015, \mnras, 451, 475
\bibitem[van Paradijs(1981)]{van81} van Paradijs, J.\ 1981, \aap, 103, 140 
\bibitem[van Paradijs(1983)]{van83} van Paradijs, J.\ 1983, in Accretion-Driven Stellar X-ray Sources,
ed.\ Lewin, W.~H.~G., \& van den Heuvel, E.~P.~J. (Cambridge: Cambridge Univ. Press), 189
\bibitem[Veledina et al.(2013)]{vel13} Veledina, A., Poutanen, J., \& Vurm, I.\ 2013, \mnras, 430, 3196
\bibitem[Veledina et al.(2015)]{vel15} Veledina, A., Revnivtsev, M.~G., Durant, M., Gandhi, P. \& Poutanen, J.\ 2015, \mnras, 454, 2855
\bibitem[Walton et al.(2012)]{wal12} Walton, D.~J., Miller, J.~M., Reis, R.~C., \& Fabian, 
A.~C.\ 2012, \mnras, 426, 473
\bibitem[Walton et al.(2013)]{wal13} Walton, D.~J., Miller, J.~M., Harrison, F.~A., et al.\ 2013, \apj, 773, L9
\bibitem[Wilms et al.(2000)]{wil00} Wilms, J., Allen, A., \& McCray, R.\ 2000, \apj, 542, 914
\bibitem[Woods et al.(1996)]{woo96} Woods, D.~T., Klein, R.~I., Castor, J.~I., McKee, C.~F., \& Bell, 
J.~B.\ 1996, \apj, 461, 767
\bibitem[Yamada et al.(2013)]{yam13} Yamada, S., Makishima, K., Done, S., et al.\ 2013, \pasj, 65,80
\bibitem[Yamaoka et al.(2001)]{yam01} Yamaoka, K., Ueda, Y., Inoue, H., et al.\ 2001, \pasj, 53, 179
\bibitem[Zdziarski \& Gierli{\'n}ski (2004)]{zdz04} Zdziarski, A.~A., \& Gierli{\'n}ski, M.\ 2004, 4, Prog. 
Theor. Phys. Suppl., 155, 99
\bibitem[Zdziarski et al.(2010)]{zdz10} Zdziarski, A.~A., Misra, R., \& Gierli{\'n}ski, M.\ 2010, \mnras, 402, 767

\end{thebibliography}
\end{document}